# On the efficacy of safety-related software standards

Mario Fusani
ISTI - Istituto di Scienza e Tecnologie dell'Informazione
Consiglio Nazionale delle Ricerche
Pisa, Italy
mario.fusani@isti.cnr.it

Giuseppe Lami
ISTI - Istituto di Scienza e Tecnologie dell'Informazione
Consiglio Nazionale delle Ricerche
Pisa, Italy
giuseppe.lami@isti.cnr.it

*Abstract*—Difficulty of safety-related software standards to help producing software for safe systems is discussed. Some research activity and other actions are proposed to focus on and possibly resolve long-lasting related problems.

## I. INTRODUCTION

In principle, safety-related software standards have a high potential for introducing valid, up-to-date research results into lifecycle processes. In fact, products resulting from these processes are admitted to public service only after conformance to standards has been demonstrated (typically in domains where authorities are established). This could be a powerful leverage for standards to give confidence to users (and to us, end-users) that innovation works for safer people and environment.

However, this big potential is somehow unexploited, as there are various reasons why, in the authors' judgment, a product declared conformant to a safety-related software standard is not necessarily, only on the basis of such statement, fit for safe use (*efficacy argument*). Here we just sketch some issues to justify this negative argument, together with possible research Actions (in italics) whose results might influence the various organizations of standards to mitigate the problems.

Our argument is two-fold: 1) a general issue regarding the never-enough-studied process-product relationship, and 2) other issues, said "stakeholder-dependent", regarding the way standards are created and used. All were pondered over various years of experience matured at the "System and Software Evaluation Center" of the ISTI (that participates in standardization activity for ISO/IEC 15504, ISO 26262, EN5012x, and assesses processes of different organizations in various Countries).

## II. GENERAL PREDICTABILITY PROBLEM

Various safety-related software standards (such as IEC 61508-3, ISO 26262, EN 50128, DO-178B) mostly prescribe requirements, also assessment requirements, for software lifecycle *processes* [1]. Then the efficacy argument can be formulated in terms of the question: can quantitative, measurable process-product relationships be known and established? Some of the mentioned standards report this as unfeasible, and the software techniques prescribed (and other techniques that support achievement of safety integrity levels) are reported as *qualitative* approaches, as opposite to *quantitative*. Much research efforts have been spent on this theme: we only refer to one short but incisive survey [4], showing that the problem can hardly be solved and ended with a series of related research questions, that are still in the same (question) status.

*Actions: improve prediction ability, both deterministic and probabilistic, with controlled experiments and analysis of collected data. Propose that lifecycle-processes-oriented, safety-related standards also cover product characteristics the way the do with processes* [1].

## III. PROBLEMS FROM STAKEHOLDERS PERSPECTIVE

More specific causes of inefficacy of standards are related to the way they are produced, the way they are adopted by their users, and the way conformance is intended by the standards themselves and by their users.

### A. Standard makers and their products

Different profiles are involved in standards Working Groups: vendors, service providers and system integrators (most of them); technology suppliers (represented); academy and consultants (some); end-users representatives (none) have different views about acceptable risks, standard usage, safety goals and conformance demonstration. As a consequence, the real purpose that emerges about standard adoption may differ from the one stated in the standard foreword.

*Actions: balance (with well pondered criteria) the composition of working groups and include end-users representatives.*

Clauses in standards are themselves requirements, but sometimes lack the very properties of system/software requirements as predicated in the standard themselves (clarity, non ambiguity, completeness, consistency).

*Actions: creation of a study group to propose "quality requirements for standards"* [2].

Management processes are little represented in most safety-related standards. They are not less necessary than technical and support processes. Having lots of good engineers in a project and poor or no management, puts the project in greater risk to fail.

*Actions:* increase *safety management; include project risks (also explaining the difference from safety risks).*

In some popular standards there is little or no guidance about how and how extensively to use Techniques and Measures (T&M) to be conformant, and in particular how to adopt Formal Methods (FM). This is left to developers and third-party evaluators to judge.

*Actions: creation of a study group, to work in T&M recommendations, that includes FM experts. In particular, show how to apply FM by letting their advantages, limits, barriers to adoption, evolution known to stakeholders [3], would also be a progress in determining conformance.*

### B. Standard users as developers

We noticed different attitudes in different roles, both with software suppliers and with system integrators.

For managers, standard conformance is scarcely perceived as a way to keep a project under control and is considered as a way to add commercial value to the product.

For engineers, the burden of producing lots of documents results in having them issued later "to show to inspectors" and anyway not as a necessary input-output among activities. As a consequence, more clerical-like than substantial conformance is pursued, this sometimes resolving in higher cost.

*Actions: foster maturity of suppliers' processes and of the procurement processes of system integrators and service providers. Reduce requests of verbose documents and explain how to compose and use documents.*

### C. Standard users as third-party evaluators

If not efficacy, is it possible to demonstrate at least conformance in an efficacious way? It seems more feasible than demonstrating fitness for safe use of a product. Such demonstration also suffers of a qualitative, and even more subjective, approach. As standards grant some liberty of interpretation in adopting software Techniques and Measures, also third party evaluators, not only developers, have their individual way to interpret conformance. This can generate or increase confusion both in suppliers and in system integrators about appointing independent evaluators.

Different kinds of third-party evaluators have different titles, including that of accredited certification bodies. Certification, whose concept and practice are not clear to all involved stakeholders, may be requested by Authorities and can give false confidence that the certified system is "fit for use", while certification is just the formal expression by the accredited body that inspections on project documentation and on the results of various test levels, executed before service, passed the (mostly qualitative) criteria set by the certification body. The fact that the certification body itself must be conformant to defined standards may increase the mentioned confidence, but the certification remains grounded on qualitative assessment results.

*Actions: Standards to clarify 3$^{rd}$ party evaluation roles, certification concept and process. Standards to indicate clearer procedures to demonstrate conformance, to shift focus toward pure (product, process) requirements than implementation aspects, and to define the role of the user stakeholders (general for B and C).*

## IV. EXAMPLES

Table I lists (respecting contractual confidentiality) examples of some of the problems found during assessments related with the issues mentioned here.

TABLE I. PROBLEMS FOUND WITH STANDARDS

| Issue | Related problems/risks |
|---|---|
| Lack of management guidance in EN 50128 (one of the reasons) | Important Euro-train project failed. Relevant financial loss. |
| Conflict of stakeholders views on using IEC 61508-3 | European big critical infrastructure control software project stuck. |
| Inter Parts – inconsistency risk of new EN 50126 (standard quality) | Standard development delayed – problems with conformance demonstration. |
| Different interpretations of Safety Integrity Level 0 (standard quality) | Systems with safety-related software made as quality-related. |
| Uncontrolled update by a vendor in a standard development (unjustified stakeholder's view prevailed) | Risk that relevant portions of the update remain without being discussed and motivated. |
| Ambiguities and incompleteness of ISO 26262-6 (standard quality) | Risk of much variability in conformance determination. |

## V. CONCLUSIONS

The examples given above cannot be more precise because of non-disclosure agreements undersigned by the assessors and also for respect among colleagues. They also don't express situations (excepting the first one) in which conformant-declared products failed meet the "fit to safe use" requirement. However, they might show evident risk of that. In the authors' intentions, the issues reported here (only part of more issues). These just support our position that a "satisfying degree of fitness for safe use" may only result from an "unplanned experiment", and that there is little hope it can be determined at system delivery. However, there *is* hope that, if the standards' quality is improved, the cost of such experiment can be greatly reduced.